\documentclass[prl,showpacs,preprint]{revtex4-1} 
\usepackage{xspace} 
\usepackage{makecell} 
\usepackage{siunitx, mhchem}
\usepackage{color,soul}
\usepackage{graphicx}
\usepackage{bm}        
\usepackage{amssymb}   
\usepackage{amsmath,mathtools} 
\usepackage{breqn} 
\usepackage{multirow}
\usepackage{array}
\usepackage{booktabs}
\usepackage[export]{adjustbox}
\usepackage{textcomp}
\usepackage[normalem]{ulem} 

\newcommand{\schrodinger}{Schr\"odinger }

\begin{document}
	
	\title{Spin transport in polarization induced two dimensional confinement of carriers in wedge shaped \textit{c}-GaN nanowalls}
	\author{Swarup Deb$^1$}
	\author{Subhabrata Dhar$^1$}
	\email{dhar@phy.iitb.ac.in}
	\affiliation{$^1$Department of Physics$,$ Indian Institute of Technology Bombay$,$ Powai$,$ Mumbai 400076$,$ India}

	\begin{abstract}
	Spin transport property of polarization induced two dimensional electron gas channel formed in the central vertical plane of a wedge-shaped \textit{c}-oriented GaN nanowall is investigated theoretically. Since the confining potential preserves the spatial symmetry between the conduction and valence band, the Rashba effect is suppressed in this system even when the shape of the wedge is asymmetric. It has been found that the relaxation of the electron spin oriented along the direction of the confinement via D'yakonov-Perel' (DP) mechanism, which is the dominant process of relaxation in this high mobility channel, is entirely switched off at low temperatures. Spin relaxation can be turned on by applying a suitable bias at the gate. Exploiting this remarkable effect, a novel all electrically driven {spin-transistor} has been proposed.     
	\end{abstract}
	
	\maketitle

In 1990, Datta and Das\cite{dattadas}  proposed a way to manipulate spin current through two dimensional electron gas (2DEG) by applying external bias at the gate. Several other proposals for a spin transistor, where spin current can be controlled by applying  a gate bias, have come up ever since\cite{david_Awschalom,PVSantos_PDI_PRB,trier_CNRS_NanoLett_electric}.  As originally proposed, narrow bandgap semiconductors with large spin-orbit coupling (SOC) remains to be the primary choice as far as spin manipulation is concerned. However, strong SOC results in large scale spin mixing of the conduction band states leading to the enhancement of spin-flip scattering rate. This makes the control of spin  a challenge in these materials\cite{jena}. Moreover,  dissipation  less flow of spin up to sufficiently long distance is an important requirement for the development of any  spin based logic  circuit.  The key approach, can be to use low SOC semiconductors as a link for spin transport and use a narrow bandgap semiconductor for controlling the  spin.\cite{jena, dykanov_book}. As a result, wide bandgap semiconductors such as GaN, ZnO with weak spin-orbit coupling  has received overwhelming attentions\cite{ZnO_Spin,GaN_spin,05_spin,04_spin,02_spin,spin_bub,exciton_SPINrelax,triangular_GaN_NW,pallabB}. Recent experimental studies demonstrate spin relaxation time of $\sim$150 ps and spin diffusion length $\sim$1\,$\mu$m at room temperature in GaN nanowires of triangular\cite{triangular_GaN_NW} and cylindrical\cite{pallabB} cross-sections.

In our earlier work, we have shown that in wedge-shaped \textit{c}-oriented GaN nanowall 2DEG can be formed in the central vertical plane of the wall as a result of the coulomb repulsion from the negative polarisation charges  developed at the two inclined facades\cite{deb}. The study further predicts a very high electron mobility in this channel\cite{deb}. This prediction is also consistent with the experimental observation of high conductivity\cite{bhasker1,bhasker3,bhasker4,bhasker5} and long phase coherence length\cite{bhasker3,ajain,Chakraborti_2018} of electrons in the networks of wedge-shaped \textit{c}-oriented GaN nanowalls. It will be interesting to understand the mechanism of spin transport through this channel. 

Eliott-Yafet (EY) and the Dyakonov-Perel (DP) are the two most prominent spin relaxation processes in semiconductors. Though EY mechanism is present both in centrosymmetric and non-centrosymmetric crystals, its effect is more prominent in the former. On the other hand, spin relaxes through DP  mechanism only in crystals lacking inversion symmetry. EY spin-flip rate enhances with the rate of momentum relaxation\cite{jena,fabian1,fabian2}. In contrast, the rate of spin relaxation through DP process decreases with the increase of momentum relaxation rate\cite{jena,fabian1,fabian2}. EY process is thus less significant in high mobility systems, where DY process effectively governs the spin relaxation. Since electron mobility in wedge-shaped \textit{c}-oriented wurtzite GaN nanowalls is expected to be significantly high and wurtzite lattice is  non-centrosymmetric,  DP mechanism is likely to dominate the  spin relaxation in this system.

Here, we have theoretically investigated the relaxation properties of electron spin in 2DEG channel formed in wedge-shaped \textit{c}-oriented GaN nanowalls\cite{deb,doi:10.1142/S2010324718400039}. Most notable finding of the study is the complete shutdown of the relaxation of spin projected along the direction of confinement. Furthermore, the phenomenon is found to be unaffected by any deviation from the symmetric shape of the wall. Interestingly however, spin relaxation can be switched on by applying a gate bias beyond a threshold point. These findings lead us to propose a novel {spin-transistor} based on this system. Novelty of this device lies in the fact that in most of the spin-transistor proposals, a control over spin relaxation is achieved through electric field induced change in the effective magnetic field experienced by the carriers in the channel\cite{ritchie_SPINFET,spin_FET}. Here, the goal can be achieved by changing the carrier concentration in the channel, which can turn DP mechanism on/off by removing/introducing  additional eigenstates below the Fermi level\cite{multibandDP,spin_bub}.

\begin{figure}[h!]
	\centering
	\includegraphics[scale=0.4]{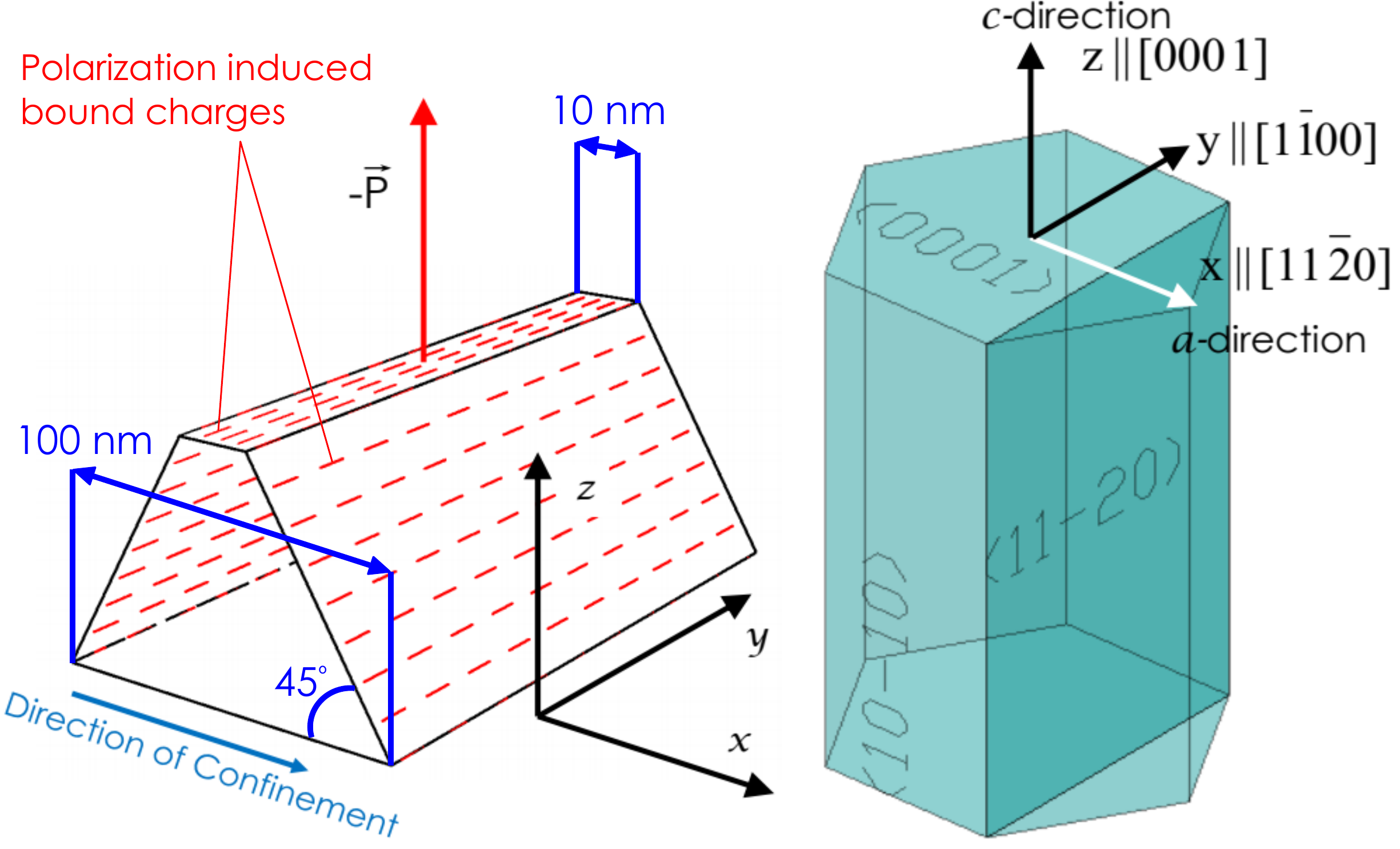}
	\caption{Schematic representation of the nanowall and the unit cell of WZ-GaN. Growth [0001] and confinement [11$\bar{2}$0] (\textit{a}-axis) directions are regarded as \textit{z} and \textit{x}-axes, respectively.}
	\label{spinfig:x-confine}
\end{figure}

Spontaneous polarization, $\vec{P}$ along $-\hat{z}$ induces a net negative polarization charge (in case of Ga-polar GaN) of density $\rho_s = \vec{P}\cdot  \hat{n}$, where $\hat{n}$ is the unit vector normal to the surface, on the inclined facades of a wedge shaped wall structure, as shown schematically in figure\,\ref{spinfig:x-confine}. In case of n-type GaN nanowalls, polarization charges on the side facades can create a repulsive force to the conduction band electrons resulting in a confinement  in the central vertical (11$\bar{2}$0) plane of the nanowall\cite{deb}. Electron movement is restricted along [11$\bar{2}$0] direction, which is regarded here as \textit{x}-axis. 

Conduction band minimum for wurtzite(WZ) GaN remains spin degenerate even after considering the effects of crystal field and SOC\cite{GaN_without_SOC}. As a result, the eigenstates, which arises due to the confinement are doubly degenerate at $k_{\shortparallel}=0$. Here, $\vec{k_{\shortparallel}}$ stands for two dimensional wavevector and can be expressed as $\vec{k_{\shortparallel}}=\vec{k_y}+\vec{k_z}$. Lack of inversion symmetry in the WZ lattice results in a $\vec{k}$ dependent spin-orbit term in the Hamiltonian, which can be expressed as\cite{01_spin,02_spin,04_spin,05_spin,spin_bub,jena}:
\begin{equation}
H_{SO}(\vec{k})=\{\alpha_R+ \beta_D (b_D k_z^2-k_x^2-k_y^2)\}(k_y \sigma_x -k_x  \sigma_y ) 
\label{eq:bulkSO}
\end{equation}
Where $\alpha_R$ determines the strength of the $k$-linear Rashba like contribution. This term arises in bulk (even in the absence of structural inversion asymmetry) as a result of the built-in electric field due to spontaneous polarization\cite{02_spin}. $\beta_D$ and $b_D$ are the Dresselhaus parameters associated with the $k^3$-terms.  In case of a 2DEG confined along [11$\bar{2}$0] direction (\textit{x}-axis), one can get an expression for $H_{SO}$ for the conduction band electrons by replacing $k_x$, $k_x^2$ terms in Eq.(\,\ref{eq:bulkSO}) by their expectation values\cite{winkler1,winkler2,chang}.
Note that  $\langle k_x \rangle=0$ for bound eigenstates. $H_{SO}$ can thus be expressed as,
\begin{equation}
H_{SO}(\vec{k})=\{\alpha_R+ \beta_D (b_D k_z^2-\langle k_x^2 \rangle-k_y^2)\}k_y \sigma_x 
\label{eq:confinedSO}
\end{equation}

$H_{SO}$ can also be expressed as: $H_{SO}(\vec{k})=\frac{\hbar}{2}\Omega (\vec{k})\cdot \vec{\sigma}$,  where $\Omega(\vec{k})$ is a vector representing the wave-vector dependent effective magnetic field and $\vec{\sigma}$ is the electron-spin. In case of bulk WZ-GaN, $\vec{\Omega}$ lies in $xy$ plane [Eq.(\,\ref{eq:bulkSO})] and its orientation is decided by the magnitude of $k_x$ and $k_y$. Interestingly, when 2DEG is confined in (11$\bar{2}$0)-plane, no matter how $\vec{k_{\shortparallel}}$ is oriented in the $yz$-plane, the effective magnetic field is always along $\hat{x}$ (+ or -) direction. However, the magnitude of the field depends upon the $y$- and $z$- components of $\vec{k_{\shortparallel}}$. Below we will see that it has a remarkable consequence on the DP spin relaxation properties of electrons confined in the (11$\bar{2}$0)-plane of the \textit{c}-oriented wedge shaped GaN nanowall.

The DP spin relaxation equation for the density, $S_i(t)$ of the spin projected along $\hat{i}$ (where, $\,i=x,y,z$) can be written as\cite{fabian2,Averkiev,long_spin_life} $\dot S_i(t)$=$-\frac{1}{2\hbar^2}\sum_{-\infty}^{\infty}\frac{\int_{0}^{\infty}d\mathcal{E}(\vec{k}_{\shortparallel})\delta f\tau_nTr([H_{-n},[H_n,\sigma_j]]\sigma_i)}{\int_{0}^{\infty}d\mathcal{E}(\vec{k}_{\shortparallel})\delta f}S_j(t)$, where $\delta f$$=$$(f_+-f_-)$, $f_{\pm}$ are the Fermi distribution functions for electrons with spin  $\pm 1/2$, $\tau_n^{-1}(k_{\shortparallel})$=$\frac{\mathcal{A}}{4\pi^2}\int_{0}^{2\pi}\mathcal{S}(\vec{k}_{\shortparallel},\vec{k}^{\prime}_{\shortparallel})[1-cos(n\theta)]d\theta$, $\mathcal{S}(\vec{k}_{\shortparallel},\vec{k}^{\prime}_{\shortparallel})$ the spin independent momentum scattering rate between $\vec{k}_{\shortparallel}$ and $\vec{k}^{\prime}_{\shortparallel}$ , $\theta$ the angle between the initial and final wave vectors, $\mathcal{A}$ the box normalization factor for the free part of the wave function of the confined electrons and $H_n$=$\int_{0}^{2\pi}\frac{d\phi}{2\pi}H_{SO}e^{-\mathtt{i} n\phi}$. It can be shown that $\dot S_x(t)=0$\cite{supp}, which implies that the DP mechanism does not alter the spin projection along $\hat{x}$ meaning the relaxation time for $x$ component of spin $\tau^s_x$ is infinite. This can also be understood from the following perspective. Since $H_{SO}$ always commutes with $\sigma_x$, $S_x$ remains a good quantum number irrespective of the direction and magnitude of $\vec{k}_{\shortparallel}$. Note that the statement is valid when all other effects which can cause a spin mixing are neglected. Our calculations further show that the relaxation times for $y$ and $z$ spin components, which follow $1/\tau^s_i$=-${\dot S_i(t)}/{S_i(t)}$ [$i=y,z$], are the same and can be expressed as\cite{supp} ${1}/{\tau^s_{y,z}}$=$\frac{8}{2\hbar^2}\left[\sum_{-1,1}  (C_1 k_{\shortparallel}+C_2 k_{\shortparallel}^3)^2\tau_n+\sum_{-3,3} C_3^2 k_{\shortparallel}^6\tau_n\right]$, where $C_1$=$(\alpha_R-\beta_D\langle k_x^2 \rangle)/2$, $C_2$=$\beta_D (b_D-3)/8$, and $C_3$=$-\beta_D (b_D+1)/8$ are material dependent constants.

\begin{figure}
	\centering
	\includegraphics[scale=0.38]{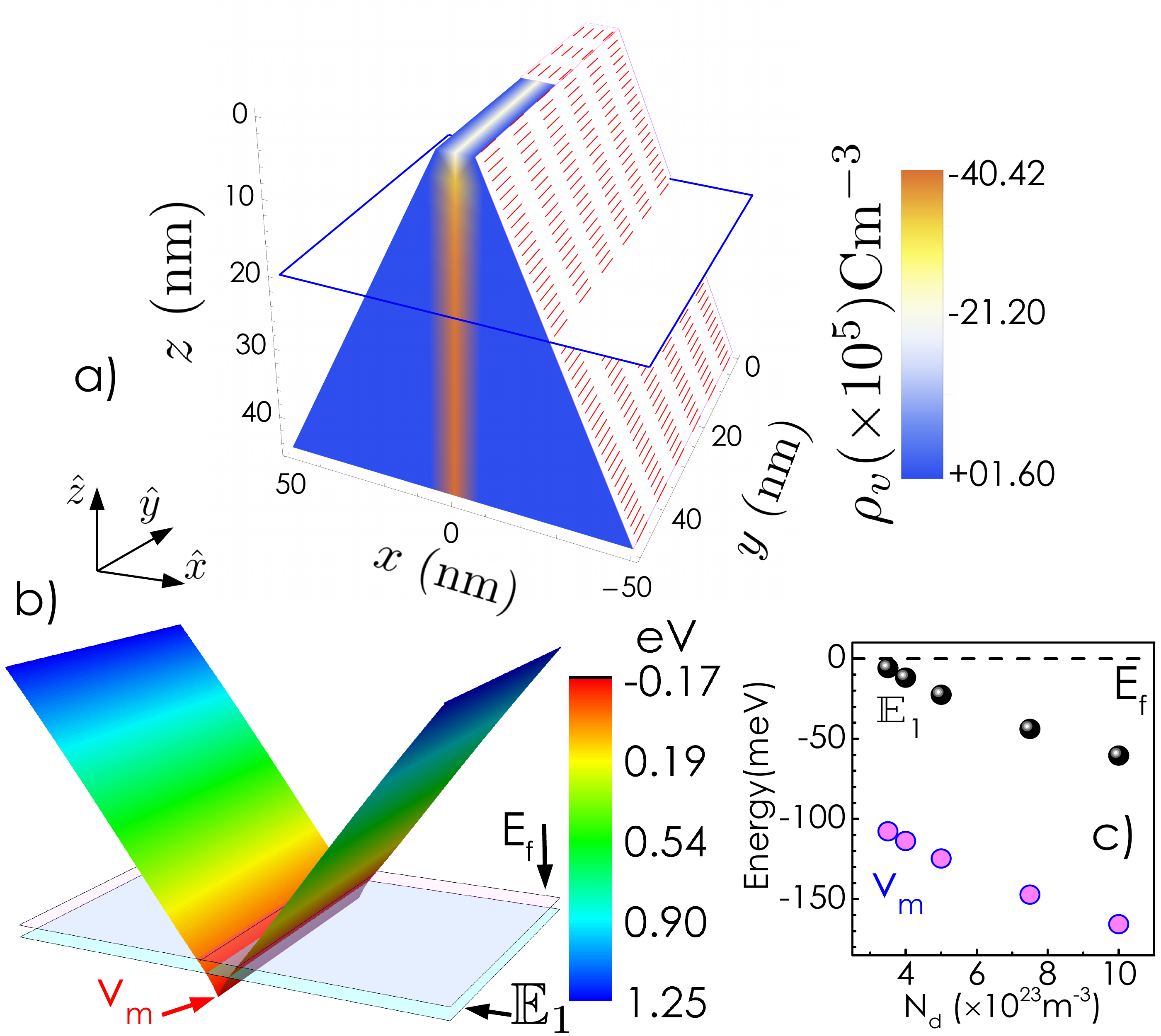}
	\caption{(a) 3D color plot for the charge density $\rho_v(x, y, z)$ inside the wedge-shaped \textit{c}-oriented WZ-GaN wall. Tip of the wall has intentionally kept half-uncovered to show the extent of charge distribution along the $y$-axis. (b) A 3D-plot for $E_c$ in $yz$-plane obtained by solving 2D-Poisson's equation. (c) Variation of the ground state energy eigenvalue and the depth of the potential well with the donor concentration,\,$N_d$.}
	\label{spinfig:2}
\end{figure}
As a test case, we have considered a wedge-shaped \textit{c}-oriented WZ-GaN nanowall with a background donor concentration ($N_d$=) of $1$$\times$$10^{24}$\,m$^{-3}$ and dimensions as shown in figure\,\ref{spinfig:x-confine}. The volumetric charge density $\rho_v(x,y,z)$ and the conduction band minimum,\,$E_c(x,y,z)$ have been obtained by solving two dimensional (2D)-Poisson's equation with appropriate boundary as well as charge neutrality conditions as described in ref\cite{deb}. Note that symmetry of the problem ensures that $\rho_v(x, y, z)$ and $E_c(x,y,z)$ are invariant along $y$-axis. $\rho_v(x,y,z)$ is shown in Fig.~\ref{spinfig:2}(a). One dimensional \schrodinger equations for $E_c(x)$ at different $z$ positions are solved to obtain  eigenfunctions and energy eigenvalues. These calculations are carried out at $T=$10\,K. Figure \ref{spinfig:2}(b) shows the conduction band profile  [$E_c(x,y)$] at a depth of 20\,nm from the tip. Evidently, the central part of the $E_c(x)$ profile goes below the Fermi surface ($E_f$), forming a trench that extends along the $y$-direction.  $\mathbb{E}_1$ denotes the first energy eigenstate of the quantum well at that depth. We have extended the calculation for several other $N_d$ values. In panel (c) the depth of the well (v$_{\text{m}}$) with respect to the Fermi energy and $\mathbb{E}_1$ obtained at $z$=20\,nm are plotted as a function of $N_d$ . Evidently, both the quantities decrease monotonically with increasing donor concentration. It should be mentioned that the range of the donor concentration is chosen in a way that only one eigenstate exists around the Fermi level, at that depth from the wall apex. Henceforth, we have shown the calculations only for the electrons lying at a depth of 20\,nm from the tip of the wall.

\begin{figure}
	\centering
	\includegraphics[scale=0.5]{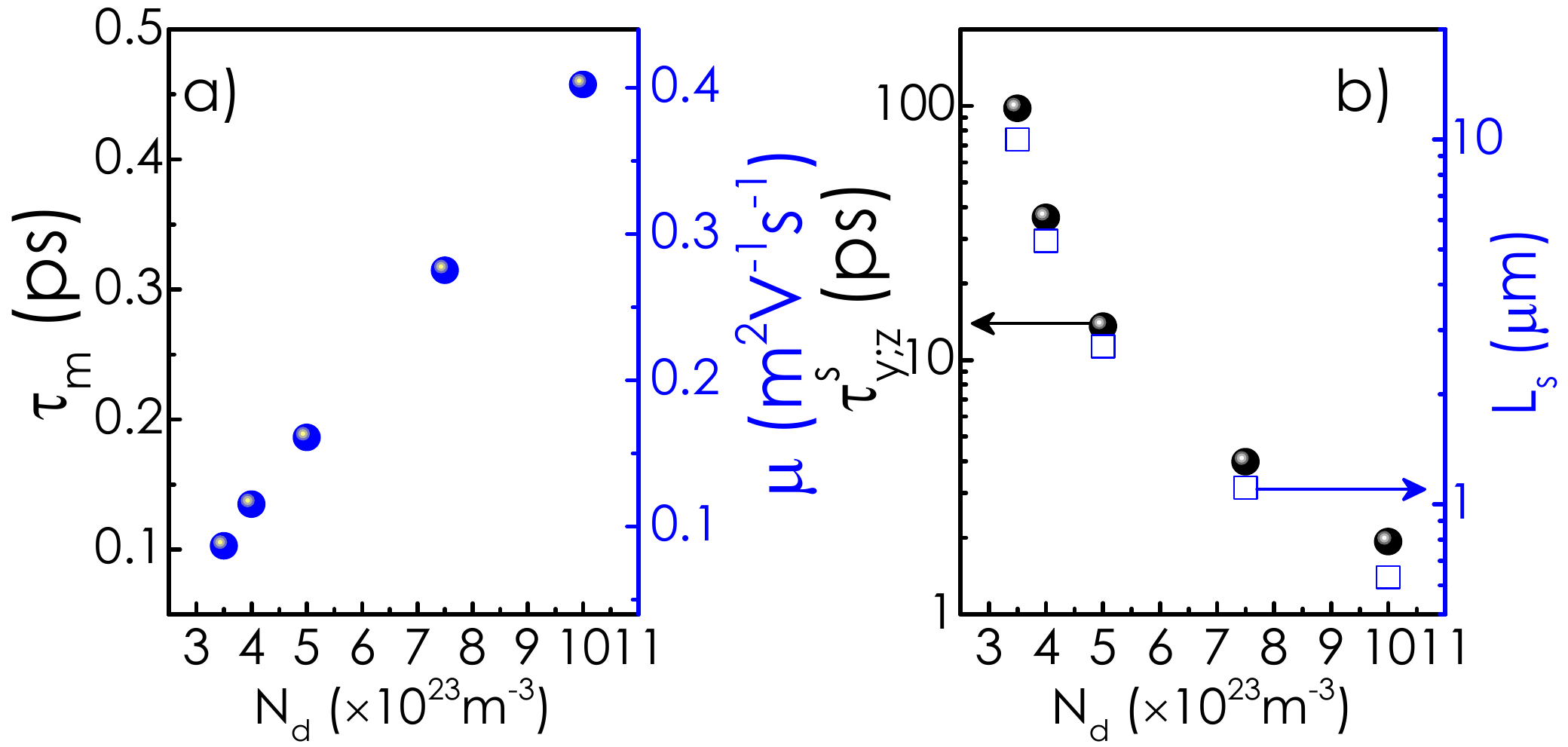}
	\caption{(a) Variation of momentum relaxation time ($\tau_m$),  mobility ($\mu$), (b) relaxation time for $y$ and $z$ component of spin($\tau^s_{y,z}$) and corresponding spin coherence length ($L_s$) as a function of $N_d$.}
	\label{spinfig:3}
\end{figure}

\begin{figure}[h]
	\centering
	\includegraphics[scale=0.51]{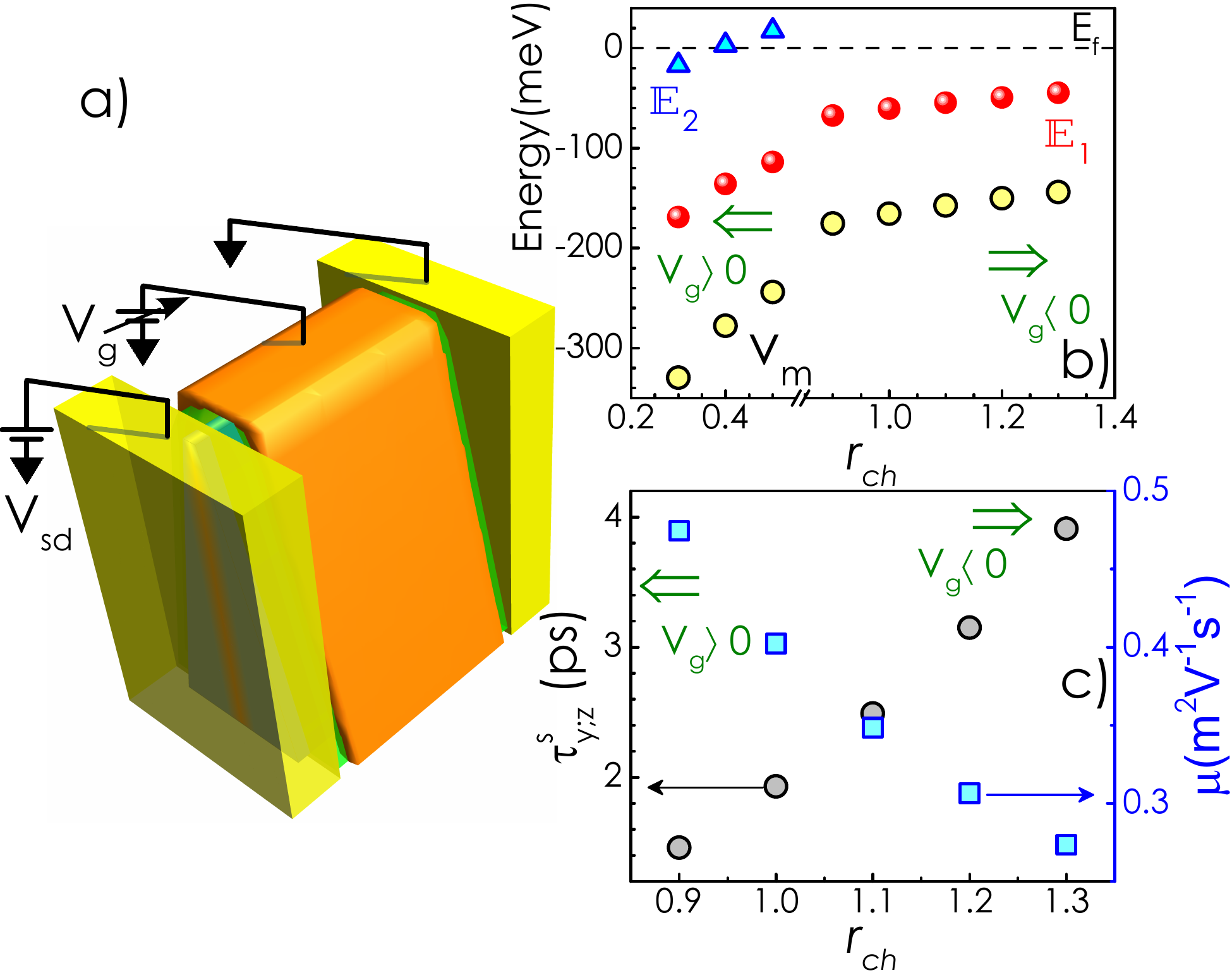}
	\caption{(a) Schematic picture of the gated device. (b) Variation of the energy eigenvalues and the depth of the potential well with $r_{ch}$. (c) Electron mobility ($\mu$) and $\tau^s_{y,z}$ as a function of $r_{ch}$.}
	\label{spinfig:4}
\end{figure}

Next, we calculate momentum relaxation time,\,$\tau_m$ of the quantum confined electrons limited by the neutral donor scattering, which plays the most significant role in deciding the electron mobility at low temperatures in this system\cite{deb}. Variation of $\tau_m$ and mobility ($\mu$) (in right $y$-ordinate) with $N_d$ is plotted in figure \ref{spinfig:3}(a), which clearly shows an increase of $\tau_m$ with the donor concentration. The effect can be attributed to the increasing separation between $\mathbb{E}_1$ and $E_f$ with  $N_d$. Increase of the separation leads to the enhancement of electron's kinetic energy, which results in the lowering of the scattering cross-section. Relaxation time for $y$ and $z$ components of spin $\tau^s_{y,z}$  as a function of $N_d$ is shown in figure \ref{spinfig:3}(b). As expected, the spin relaxation time decreases as $\tau_m$ increases. It should be noted that $\tau^s_{y,z}$ comes out to be $\sim$100\,ps for the nanowall with $N_d=0.35\times10^{18}$\,cm$^{-3}$. Interestingly, a few factor change in donor density alters the spin relaxation time  by about two orders of magnitude. The spin coherence length, $L_s$  = $\tau^s v_f$, where $v_f$ is the Fermi velocity,  is also plotted as a function of $N_d$  in the same panel. Note that $L_s$ for the lowest donor concentration comes out to be as high as 10\,$\mu$m.  As obtained earlier, the spin coherence time ($\tau^s_x$) for the spin projected along $x$-direction is infinite as far as DP mechanism is concerned. Relaxation of the spin projected along $x$-direction is thus governed mainly by EY mechanism. One can estimate EY spin relaxation time from $\tau_m$, which comes out to be of the order of a few $\mu$s\cite{optical_orientation, bubPRB}(though the relation is strictly valid for cubic GaN). 
One way to manipulate spin transport in this system is to control the carrier concentration in the channel through gate bias. The idea is that with increasing carrier concentration, kinetic energy of the electrons around the Fermi level increases. This, in turn, can change both $\mu$ the electron mobility and $\tau^s_i$ the spin relaxation time. In order to calculate these changes, one needs to incorporate the effect of gate voltage in the solution of Poisson equation. Gate contact and the semiconducting channel together form a capacitor [see figure\,\ref{spinfig:4}(a)]. When source and drain electrodes are grounded and a positive(negative) gate voltage is applied, some amount of electrons are pumped(removed) into(from) the channel by the power supply. Since the semiconductor is  no longer charge neutral, the Poisson's equation has to be solved by satisfying appropriate positive to negative charge ratio condition (instead of satisfying charge neutrality) to obtain the $E_c$ profiles. Total positive  to negative charge ratio ($r_{ch}$) should be less(greater) than 1 when sufficiently positive(negative) gate voltages are applied.  As shown in Fig.\,\ref{spinfig:4}(b), the gap between $\mathbb{E}_1$ and $E_f$ decreases as $r_{ch}$ increases. Note that when $r_{ch}$ is sufficiently less than unity [high $(+)ve$ gate voltages], more than one bound states are formed below the Fermi level. 
Variation of $\mu$ and $\tau^s_{y,z}$ with $r_{ch}$ are shown in Fig.\,\ref{spinfig:4}(c). As the gap between $\mathbb{E}$ and $E_f$ decreases with the increase of $r_{ch}$, $\mu$ reduces while $\tau^s_{y,z}$ enhances. 

\begin{figure}
	\centering
	\includegraphics[scale=0.56]{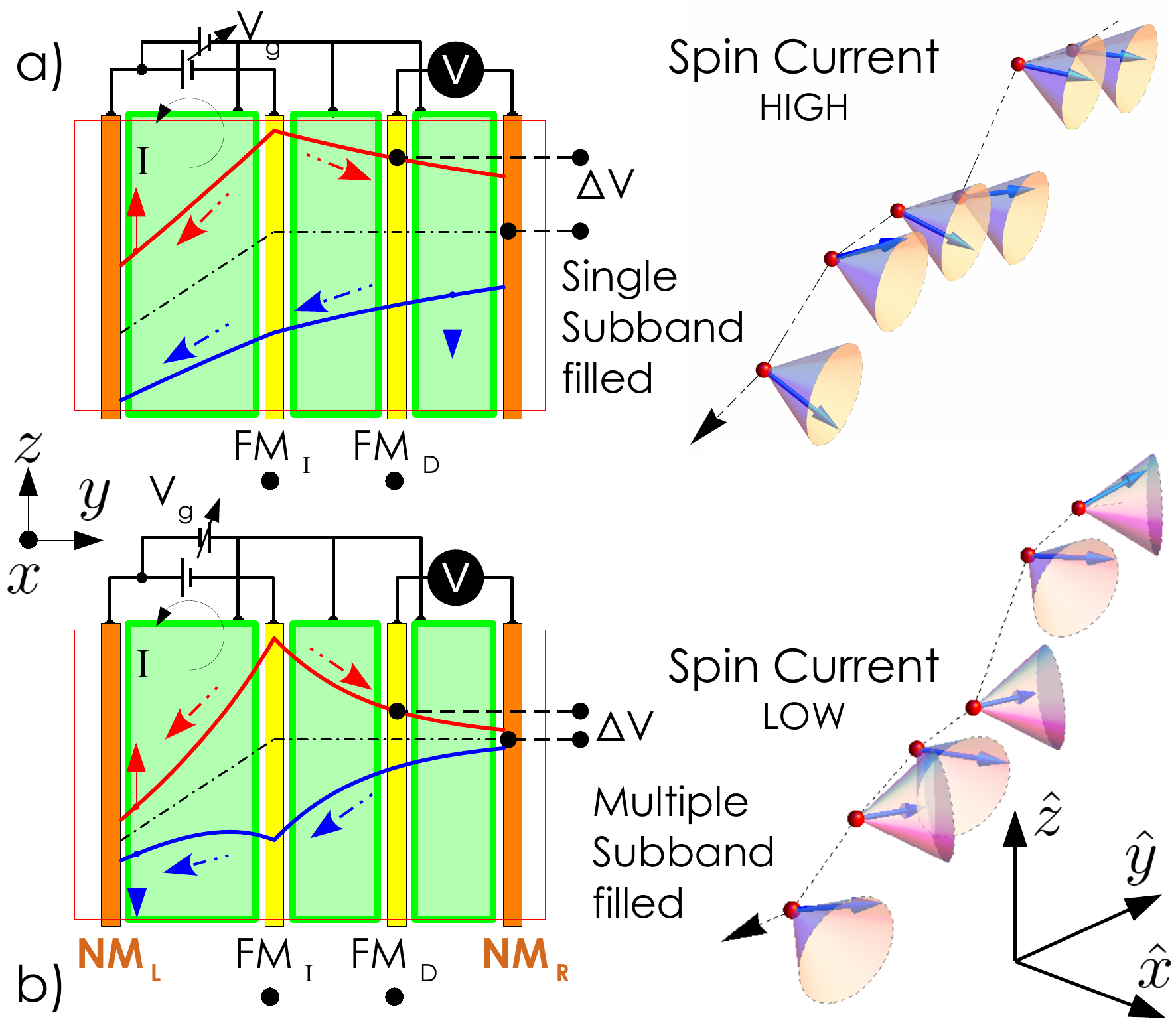}
	\caption{Schematic depiction of the variation of the spin-up (red curve), spin-down (blue curve), and average (dashed black curve) chemical potentials across the channel of {the proposed spin-transistor}. Broken arrows show the component of spin-resolved charge current. Red(blue) solid arrows represent +(-)\textit{x} spin states. The green pads stand for the gate electrodes. When a single subband is filled, electron spin always precesses about $\hat{x}$ (+ or -) direction irrespective of any change in the direction of $\vec{k_{\shortparallel}}$ due to scattering. On the other hand, when multiple subbands are filled, each scattering event results in a change in $\vec{k_{\shortparallel}}$ that alters the direction of effective magnetic field and hence the spin precession axis.}
	\label{spinfig:5}
\end{figure}

As shown earlier, spin component along \textit{x}-direction does not relax in this system as DP process is completely switched-off for this spin projection.  This assertion is strictly valid when only a single subband is occupied. When more than one eigenstates are formed below the Fermi level, DP relaxation is activated even for the \textit{x} projection of spin as in this case, wavefunction for the electrons near the Fermi level becomes a linear superposition of multiple eigenstates. In such a situation, $\langle k_x \rangle$ is no longer equal to zero\cite{multibandDP}. Finite value of  $\langle k_x \rangle$ results in a non-zero $y$ component of $\vec\Omega$. Thus, $H_{SO}$ does not commute with any of the spin components resulting in a finite relaxation time for the spin density $S_x$ as well. This property can be exploited to envisage a novel type of {spin-transistor}. A schematic of such a device is shown in figure\,\ref{spinfig:5}(a). In this nonlocal spin valve {(spin-transistor)} device, the pair of contacts in the middle are made of ferromagnetic metals (FM), while the contacts on edges are nonmagnetic (NM). The FM contact on left is the spin injector (FM$_{\text{I}}$) while the other works as the detector (FM$_{\text{D}}$). Consider the case when both FM$_{\text{I}}$ and FM$_{\text{D}}$ are coupled to spin up (with respect to \textit{x} axis) electrons. At an adequately high negative gate voltage, when only one subband is filled, the \textit{x}-polarized spin up current injected from FM$_{\text{I}}$ reaches the detector electrode without losing its coherence. As a result, the spin current through the channel is high and `spin-resolved charge voltage' measured between FM$_{\text{D}}$ and NM$_{\text{R}}$ will be sufficiently high. Upon application of a large positive gate bias, more than one subbands are filled.  At this condition, the direction of the effective magnetic field ($\vec{\Omega}$), which is the Larmor precession axis of electron spin, get randomized due to successive scattering events. As a result, the \textit{x} projection of spin also relaxes. This should result in a rapid decay of pure spin current in the channel between FM$_{\text{I}}$ and FM$_{\text{D}}$, as depicted in figure\,\ref{spinfig:5}(b). The device can thus act as a {spin-transistor}. Note that the nature of quantum confinement in this system is such that it preserves the symmetry between conduction and valence band everywhere irrespective of the fact whether the nanowedge is geometrically symmetric or not. This results in the suppression of the Rashba effect, which comes as an added advantage in maintaining spin coherence of electrons in this system. Note that, this novel 2DEG system is substantially different from  the 2DEG, formed at the heterojunction of \textit{a}-directional GaN/AlGaN as discussed in supplementary information\cite{supp}.

Spin relaxation of electrons in a 2DEG formed at the central vertical plane of a \textit{c}-oriented wedge-shaped WZ-GaN nanowall is theoretically investigated. It has been found that the component of spin projected  in the plane of the confinement relaxes  through DP mechanism with a time-scale of a few tens of picoseconds everywhere in the channel, while the spin component along the direction of confinement never relaxes through DP process in a part of  the channel, where the Fermi-level occupies only one subband. However, by applying appropriate positive gate bias, electron concentration in the well can be sufficiently enhanced, which can bring in more than one  subbands below the Fermi-level in most of the channel. At this situation, DP relaxation mechanism is switched-on also for the spin component along the confinement direction. One can envisage  a novel  {spin-transistor}, where this property can be exploited to electrically manipulate spin current.

We acknowledge the financial support by Department of Science and Technology, Government of India under the project Code:CRG/2018/001343.

\bibliographystyle{unsrt}
\bibliography{Reference}
	
\noindent
{\huge Supplementary Information}
\\

\noindent
\textbf{{\Large{Spin transport in polarization induced two dimensional confinement of carriers in wedge shaped \textit{c}-GaN nanowalls}}}
\\
{\large Swarup Deb, Subhabrata Dhar}\\
Department of Physics, Indian Institute of Technology Bombay, Powai, Mumbai-400076, India
\\
\vspace{1cm}
\\
\noindent
\textbf{{\large{S1. Derivation of spin relaxation time}}}

In order to calculate the DP spin relaxation time we follow the formalism described by N. S. Averkiev \textit{et al.}\cite{Averkiev} and J. Fabian \textit{et al.}\cite{fabian2}.
The DP relaxation rate for spin density,\,$S_i(t)$ projected along $\hat{i}$ ($\,i=x,y,z$) can be written as\cite{long_spin_life}:
\begin{equation}
	\dot S_i(t)=-\frac{1}{2\hbar^2}\sum_{n=-\infty}^{n=\infty}\frac{\int_{0}^{\infty}d\mathcal{E}(\vec{k}_{\shortparallel})\delta f\tau_nTr([H_{-n},[H_n,\sigma_j]]\sigma_i)}{\int_{0}^{\infty}d\mathcal{E}(\vec{k}_{\shortparallel})\delta f}S_j(t)
	\label{eq:dprelaxationtime}
\end{equation}
where, $\delta f$$=$$(f_+-f_-)$, $f_{\pm}$ are the Fermi distribution functions for electrons with spin  $\pm 1/2$. $\tau_n$ are the relaxation times\cite{DP_vs_T,long_spin_life}, which can be expressed as:
\begin{equation}
	\tau_n^{-1}(k_{\shortparallel})=\frac{\mathcal{A}}{4\pi^2}\int_{0}^{2\pi}\mathcal{S}(\vec{k}_{\shortparallel},\vec{k}^{\prime}_{\shortparallel})[1-cos(n\theta)]d\theta
\end{equation}
where, $\mathcal{S}(\vec{k}_{\shortparallel},\vec{k}^{\prime}_{\shortparallel})$  the spin independent momentum scattering rate between $\vec{k}_{\shortparallel}$ and $\vec{k}^{\prime}_{\shortparallel}$ , $\theta$ the angle between the initial and final wave vectors, $\mathcal{A}$ the box normalization factor for the free part of the wave function of the confined electrons. $H_n$ are the Fourier harmonics of spin-orbit Hamiltonian:
\begin{equation}
	H_n=\int_{0}^{2\pi}\frac{d\phi}{2\pi}H_{SO}e^{-\mathtt{i} n\phi}
\end{equation}
Because of the presence of linear and cubic terms of $k_{\shortparallel}$ in $H_{SO}$, $H_n$ survives only for $n=\pm 3\text{ and }\pm 1$.

 It can be shown that:
\begin{eqnarray}
H_{+1}=H_{-1}&=&(\alpha_R-\beta_D\langle k_x^2 \rangle)/2\,k\sigma_x+\beta_Db_D/8\, k^3\sigma_x-3\beta_D/8\,k^3\sigma_x\nonumber\\
&=&C_1 k \sigma_x+C_2 k^3 \sigma_x\\
H_{+3}=H_{-3}&=&-\beta_Db_D/8\, k^3\sigma_x-\beta_D/8\,k^3\sigma_x\nonumber\\
&=&C_3 k^3 \sigma_x
\end{eqnarray}

For, $\dot S_x(t)$ the trace, $Tr(\cdots)$ in equation\,(\ref{eq:dprelaxationtime}) takes the form:
\begin{eqnarray}
Tr([H_{-n},[H_n,\sigma_j]]\sigma_x)\propto Tr([\sigma_x,[\sigma_x,\sigma_j]]\sigma_x);\, j=x,y,z
\label{eq:trace}
\end{eqnarray}
In the above relation, the $k_{\shortparallel}$ dependent prefactors are dropped for simplicity. We have also taken the advantage of the fact that in such quantum confined system $H_{SO}$ get coupled to $\sigma_x$, only. The Pauli matrices follow the commutation relation $[\sigma_i,\sigma_j]=2\mathtt{i}\xi_{ijk}\sigma_k$, $\xi_{ijk}$ is the Levi-Civita tensor. Using this relation the trace can be evaluated as:
\begin{eqnarray}
Tr([\sigma_x,[\sigma_x,\sigma_j]]\sigma_x)&=&2\mathtt{i}Tr([\sigma_x,\xi_{xjk}\sigma_k]\sigma_x)\nonumber\\
&=&-4\xi_{xjk}\xi_{xkj}Tr(\sigma_j\sigma_x)
\end{eqnarray}
Again, $Tr(\sigma_j\sigma_x)\neq0$ only if $j=x$ but in that case $\xi_{xjk}$ or $\xi_{xkj}$ are 0 and vice versa. Thus the $Tr(\cdots)$ term in the numerator of equation\,(\ref{eq:dprelaxationtime}) becomes 0 for $\dot S_x(t)$, leading to 
\begin{equation}
\dot S_x(t)=0
\end{equation}

Derivation of $\dot S_y$ and $\dot S_z$ is slightly tedious. The trace in the numerator of equation\,(\ref{eq:dprelaxationtime}) has to be evaluated for all possible combination of $i$, $j$ and $n$. As an example we evaluate the terms for $\dot S_y$ in table\,\ref{tab2}. Thus $i$ has been replaced with $y$.


\begin{center}
	\begin{table}[h]
		\centering
		\caption{Evaluating ``$\tau_nTr([H_{-n},[H_n,\sigma_j]]\sigma_i)S_j(t)$'' from equation\,(\ref{eq:dprelaxationtime}) as $i=y$}
		{\small{
				\begin{tabular}{c|l|l|l} 
					\textit{n} & \multicolumn{1}{c|}{\textit{j=x}} & \multicolumn{1}{c|}{\textit{j=y}} & \multicolumn{1}{c}{\textit{j=z}} \\
					\hline
					\hline
					+3 & $\tau_3Tr([H_{-3},[H_3,\sigma_x]]\sigma_y)S_x(t)$  & $\tau_3Tr([H_{-3},[H_3,\sigma_y]]\sigma_y)S_y(t)$  & $\tau_3Tr([H_{-3},[H_3,\sigma_z]]\sigma_y)S_z(t)$  \\[0.5ex]
					& = 0 & = $\tau_3 8 C_3^3k^6 S_y(t)$ & = 0\\[1ex]
					
					+1 & $\tau_1Tr([H_{-1},[H_1,\sigma_x]]\sigma_y)S_x(t)$  & $\tau_1Tr([H_{-1},[H_1,\sigma_y]]\sigma_y)S_y(t)$  & $\tau_1Tr([H_{-1},[H_1,\sigma_z]]\sigma_y)S_z(t)$ \\[0.5ex]
					& = 0 & = $\tau_1 8(C_1 k +C_2 k^3)^2  S_y(t)$ & = 0\\[1ex]

					-1 & $\tau_{-1}Tr([H_{1},[H_-1,\sigma_x]]\sigma_y)S_x(t)$  & $\tau_{-1}Tr([H_{1},[H_{-1},\sigma_y]]\sigma_y)S_y(t)$  & $\tau_{-1}Tr([H_{1},[H_{-1},\sigma_z]]\sigma_y)S_z(t)$\\[0.5ex]
					& = 0 & = $\tau_{-1} 8(C_1 k +C_2 k^3)^2  S_y(t)$ & = 0\\[1ex]
					
					-3 & $\tau_{-3}Tr([H_{3},[H_-3,\sigma_x]]\sigma_y)S_x(t)$  & $\tau_{-3}Tr([H_{3},[H_{-3},\sigma_y]]\sigma_y)S_y(t)$  & $\tau_{-3}Tr([H_{3},[H_{-3},\sigma_z]]\sigma_y)S_z(t)$ \\[0.5ex]
					& = 0 & = $\tau_{-3} 8 C_3^3k^6 S_y(t)$ & = 0
				\end{tabular}
		}}
		\label{tab2}
	\end{table} 
\end{center}
$\dot S_z$ can also be evaluated in a similar fashion. It is important to note that $\dot S_y$ only depends on the $y$ spin projection, $S_y$. The same hold for the \textit{z} component also. Thus one can write the spin relaxation time of \textit{y} and \textit{z} components as:
\begin{equation}
\frac{\dot S_y(t)}{S_y(t)}=-\frac{1}{\tau^s_y}\text{; and }\frac{\dot S_z(t)}{S_z(t)}=-\frac{1}{\tau^s_z}
\end{equation}
Following reference\cite{Averkiev}, we write $\tau^s_y$ and $\tau^s_z$ as
\begin{equation}
\frac{1}{\tau^s_y}=\frac{1}{\tau^s_z}=\frac{1}{\tau^s_{y,z}}=\frac{1}{2\hbar^2}\left[\sum_{n=-1,1}8 (C_1 k_{\shortparallel}+C_2 k_{\shortparallel}^3)^2\tau_n+\sum_{n=-3,3}8 C_3^2 k_{\shortparallel}^6\tau_n\right]
\label{eq:final_tau}
\end{equation}
Following parameter values are used during numerical calculation: $\alpha_R$=9.0\,meV\AA{}; $b_D$=3.959; $\beta_D$=0.32\,eV\AA{}$^3$\cite{04_spin}
\vspace{1cm}
\\
\noindent
\textbf{{\large{S2. Spin dynamics:\textit{a-}plane 2DEG in a AlGaN/GaN Heterojunction versus \textit{a-}plane 2DEG in $c$-oriented GaN nanowedge  }}}

At a first glance, 2D electron gas confined in \textit{a-}plane of AlGaN/GaN\cite{a-AlGaN/GaN} appears very similar to the present case of the 2DEG formed in a $c$-oriented wedge shaped GaN nanowall structure. However, unlike the present case, such a heterojunction lacks spatial inversion symmetry(\textit{SIA}) leading to the existence of a net electric field. This assertion requires a more careful discussion.

In a quantum confined system, the effective electric field can be quantified by the expectation value of the field \textit{i.e.} $-\langle \frac{\partial E_c}{\partial x} \rangle/q_e$. Where, $E_c$ is the conduction band energy profile in real space and $q_e$ is the electron charge. Using the identities from Ehrenfest theorem:
\begin{eqnarray}
\langle - \frac{\partial E_c}{\partial x} \rangle&=&\frac{d}{dt}\langle p \rangle\\
&=&\frac{d}{dt}\left(m\frac{d}{dt}\langle x \rangle\right)
\end{eqnarray}
For a bound eigen state, time derivative of $\langle x \rangle$ = 0 and hence the effective electric field is also zero. This was first pointed out by T. Ando\cite{ando_rashba_00,ando_rashba_01} who argued that the spin splitting of conduction band must be negligibly small in such system. Later, Lassnig\cite{lassnig} showed that Rashba field in such quantum well system will be proportional to the average electric field of valence band seen by the conduction band electrons \textit{i.e.} $\int |\psi_c|^2 \frac{\partial E_v}{\partial x}dx$. Where, $\psi_c$ denotes the wave-function for the electrons at the conduction band edge and $E_v$ is the valence band edge profile. In a \textit{a-}plane 2DEG of AlGaN/GaN heterojunction, the band offset breaks the symmetry between the conduction and valence band across the junction. As a result, $\int |\psi_c|^2 \frac{\partial E_v}{\partial x}dx$ does not vanish for a bound state while $\int |\psi_c|^2 \frac{\partial E_c}{\partial x}dx$ does. Due to Rashba effect, this average field gives rise to an effective magnetic field acting perpendicular to $\hat{x}$\cite{VB_Rashba1,ando_rashba_01} and $H_{SO}$ becomes
\begin{equation}
H_{SO}(\vec{k})=\{\alpha_R+ \beta_D (b_D k_z^2-\langle k_x^2 \rangle-k_y^2)\} \left(
\begin{array}{ccc}
k_y  & 0  & 0  \\
\end{array}
\right)\left(
\begin{array}{c}
\sigma_x\\
\sigma_y\\
\sigma_z
\end{array}
\right)
+\alpha_{x}\left(
\begin{array}{ccc}
0  & k_z  & -k_y  \\
\end{array}
\right)\left(
\begin{array}{c}
\sigma_x\\
\sigma_y\\
\sigma_z
\end{array}
\right)
\label{eq:a-confinedSO}
\end{equation}
Clearly, $H_{SO}$ does not commute with any of the $\sigma$'s. Spin density $S$ thus follows a different dynamics than in the present case. It should be noted that in case of the \textit{a-}plane 2DEG in \textit{c}-oriented wedge shaped GaN nanowalls, both $\langle \psi_c|\frac{\partial E_c}{\partial x}|\psi_c\rangle$=$\langle \psi_c|\frac{\partial E_v}{\partial x}|\psi_c\rangle$=0 as the symmetry between the conduction and valence band across the junction is intact. This is true even when the shape of the wedge is asymmetric. Therefore, the spin dynamics remains to be the same for both asymmetric and symmetric nanowedges.
\vspace{1.0cm}
\\
\noindent



\end{document}